\documentclass[conference]{IEEEtran}
\IEEEoverridecommandlockouts

\usepackage{epsfig,rotating,setspace,latexsym,amsmath,epsf,amssymb,amsfonts,bm,theorem,cite,enumerate,longtable,accents,float,physics,colortbl,array,algorithm,algorithmic,graphicx,epsf,authblk,epstopdf,url,xcolor, soul,multirow,bbm,mathtools,comment}
\usepackage[center]{qtree}
\usepackage{tree-dvips}
\usepackage[linguistics]{forest }
\usepackage{diagbox}
\usepackage[utf8]{inputenc} 
\usepackage[T1]{fontenc}
\usepackage{ifthen}
\usepackage{cite}

\newtheorem{theorem}{Theorem}
\newtheorem{corollary}{Corollary}

\newtheorem{remark}{Remark}

\newenvironment{Proof}[1]{\medskip\par\noindent{\bf Proof:\,}\,#1}{{\mbox{\,$\blacksquare$}\par}}

\newcommand{\cq}{{\mathcal{Q}}}

\begin{document}

\allowdisplaybreaks

\title{Storage-Rate Trade-off in A-XPIR}
\author{Mohamed Nomeir \qquad  Sennur Ulukus\\
	\normalsize Department of Electrical and Computer Engineering\\
	\normalsize University of Maryland, College Park, MD 20742 \\
	\normalsize \emph{mnomeir@umd.edu} \qquad \emph{ulukus@umd.edu}}

\maketitle

\begin{abstract}
We consider the storage problem in an asymmetric $X$-secure private information retrieval (A-XPIR) setting. The A-XPIR setting considers the $X$-secure PIR problem (XPIR) when a given arbitrary set of servers is communicating. We focus on the trade-off region between the average storage at the servers and the average download cost. In the case of $N=4$ servers and two non-overlapping sets of communicating servers with $K=2$ messages, we characterize the achievable region and show that the three main inequalities compared to the no-security case collapse to two inequalities in the asymmetric security case. In the general case, we derive bounds that need to be satisfied for the general achievable region for an arbitrary number of servers and messages. In addition, we provide the storage and retrieval scheme for the case of $N=4$ servers with $K=2$ messages and two non-overlapping sets of communicating servers, such that the messages are not replicated (in the sense of a coded version of each symbol) and at the same time achieve the optimal achievable rate for the case of replication. Finally, we derive the exact capacity for the case of asymmetric security and asymmetric collusion for $N=4$ servers, with the communication links $\{1,2\}$ and $\{3,4\}$, which splits the servers into two groups, i.e., $g=2$, and with the collusion links $\{1,3\}$, $\{2,4\}$, as $C=\frac{1}{3}$. More generally, we derive a capacity result for a certain family of asymmetric collusion and asymmetric security cases. 
\end{abstract}

\section{Introduction}
Private information retrieval (PIR) was introduced in \cite{chor}. In PIR, a user wants to retrieve one out of $K$ messages, from $N$ non-colluding servers, without revealing the index of the retrieved message. In \cite{c_pir}, the capacity of PIR, i.e., the highest possible ratio between the required message length and the number of total downloaded symbols, was derived for the case of replicated storage, i.e., when all the servers store the same set of messages. In \cite{colluding}, the capacity with $T$-colluding servers, where any $T$ servers can collude and exchange user queries, was derived in the fully replicated setting. The problem for $T$-colluding PIR was further investigated in \cite{arbitrarycollusion}, where the collusion is arbitrary but known by the user, and the capacity was derived for the fully replicated storage. Similarly, the problem with $E$-eavesdroppers was analyzed in \cite{sun_eaves}, where any $E$ of the links may be eavesdropped, and the case of arbitrary but known eavesdropping pattern was analyzed in \cite{nan_eaves}. In the setting where the stored messages must be secure against any $X$-communicating servers, coined as $X$-secure PIR (XPIR), a scheme was developed in \cite{csa}, however, a closed-form for the capacity is still an open problem. Similarly, \cite{asymmetric} analyzed the asymmetric XPIR case with an arbitrary but given structure of $X$-communicating databases with symmetric $T$-collusion (A-XTPIR). The capacities for some specific cases were derived. In addition, a scheme was developed that works in any given arbitrary A-XTPIR setting. 

The full replication requirement is becoming very stringent due to the ever-increasing data volume. To resolve this issue, some papers investigated the capacity of specific replication patterns \cite{grpahbased_pir, shreya_graph_1, shreya_graph_2, shreya_graph_3, shreya_graph_4, JAFAR_STAR_4}. This approach is coined as graph-based PIR since a graph represents the partial replication structure. For instance, in \cite{grpahbased_pir, shreya_graph_1, shreya_graph_2, shreya_graph_3, shreya_graph_4, JAFAR_STAR_4}, the nodes represent the servers and the hyper-edges represent the shared messages among the servers. Conversely, in \cite{karim_graph}, the nodes represent the messages and the hyper-edges represent the servers that store the message nodes in their storage. In addition, \cite{jafar_asymptotic_graph} analyzed the capacity of arbitrary graphs for asymptotically-many messages. An alternative approach to PIR from the storage perspective was investigated in \cite{chao_storage, sun-chao-storage, multiround_pir}, where the trade-off between the optimal download rate and the average storage was analyzed. Thus, the goal there became finding the capacity region for the feasible storage-rate pairs instead of finding the optimal rate for a certain storage pattern. 

In this paper, we focus on the case of A-XPIR with no collusion, i.e., $T=1$. We derive the achievable region for the case with $N=4$ servers and two communication links. We show, in this case, that the storage can be minimized while having the same optimal rate for the fully replicated setting. Interestingly, the approach adopted allows some of the servers to collude without affecting the rate or the storage. Thus, we derive the capacity result for this case. In addition, we derive bounds for the general setting. 

The rest of the paper is organized as follows. Section \ref{sec_formulation} provides the system model and formulates the requirements. Section \ref{sec_main_results} provides the main theorems and their proofs, along with the trade-off graph for the case of $N=4$ servers, two non-overlapping sets of communicating servers, and $K=2$ messages. Section \ref{sec_example} provides an example that achieves the optimal download rate for the replicated servers with a lower storage overhead; in addition, it gives two important capacity results for certain families of arbitrary collusion and security.

\section{Problem Formulation}\label{sec_formulation}
\begin{figure}[t]
      \centering
      \includegraphics[width=0.6\columnwidth]{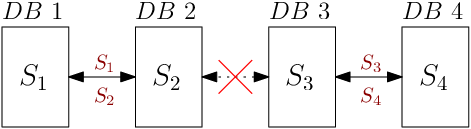}
      \caption{System model with $4$ servers where servers $1$ and $2$ communicate, and servers $3$ and $4$ communicate. No other links between servers exist, e.g., servers $2$ and $3$ do not communicate.}
      \label{fig_formulation}
\end{figure}

Consider a system with $N$ databases, $K$ messages, $W_{[K]}$ generated uniformly at random from $\mathbb{F}_q ^ L$. In addition, assume that the system must at least be secure against any of the databases, i.e., no single server can know any information about the message symbols. Let the $M$ communication links be defined using the communication matrix $B_X$ with dimensions $N \times M$, where if $i_1$, $i_2$, and $i_3$ databases communicate and $j_1$, $j_2$, and $j_3$ communicate then $B_X(i_1,1)=B_X(i_2,1)=B_X(i_3,1)=1$ and $B_X(j_1,2)=B_X(j_2,2)=B_X(j_3,2)=1$. Define the similarity matrix $\Lambda$ as an $\lceil \frac{N}{2}\rceil \times N$ matrix, where $\Lambda(i,k)=\Lambda(i,j)=1$ means that the $k$th and $j$th servers are in the same group. Thus, the optimization problem for grouping databases is, as defined in \cite{asymmetric}, given as follows
\begin{align}\label{main_opt}
    \max_{S} & \quad \sum_{i} \mathbbm{1}\{\Lambda_i \neq \mathbf{0}_{1\times N}\}\nonumber\\
    \text{s.t.} &\quad \Lambda B^c_{X} \geq \mathbf{1}_{\frac{N}{2} \times M}, \nonumber\\
    &\quad \Lambda(i,j) \in \{0,1\}, \quad  j=[N],\quad i=\left[\left\lceil \frac{N}{2}\right\rceil\right], \nonumber\\
    &\quad \sum_{j=1}^{N} \Lambda(i,j) \neq 1, \quad i=\left[\left\lceil \frac{N}{2}\right\rceil\right], \nonumber\\
    &\quad \Lambda_i^T \Lambda_j = 0, \quad i\neq j,
\end{align}
where $\Lambda_i$ is the $i$th row in $\Lambda$, $\mathbf{0}_{a \times b}$ is the all zeros $a \times b$ matrix, similarly $\mathbf{1}_{a \times b}$ is the all ones matrix, and $B^c_X$ is the binary complement of $B_X$. Fig.~\ref{fig_formulation} shows an example of our system model when servers 1 and 2 communicate and servers 3 and 4 communicate, with no other communication links.

\begin{remark}
    Recall that for $B_X$, we remove as much redundancy as possible, i.e., if servers 1, 2, and 3 communicate then we include only this communication link in $B_X$ without adding a link that shows that server 1 communicates with server 2, and server 2 communicates with server 3, etc., individually. However, the other way around is not the same. For example, if we have server 1 communicating with server 2, server 2 communicating with server 3, and server 3 communicating with server 1, we add them as three separate links in the communication matrix $B_X$. We consider these two cases as two separate scenarios, where in the first case each of the servers can keep $3KL$ symbols in their storage individually at the same time, thus we need to protect the symbols from three communicating servers. In the latter, we consider that each server can store $2KL$ symbols, thus they cannot store the three collectively together at the same time, i.e., they can store one other server's data along with their own only. Thus, in the latter, we have to protect from two-server communications.    
\end{remark}

\begin{remark}
    We intuitively explain the optimization problem defined in \eqref{main_opt} as was also done in \cite{asymmetric}. The goal of the optimization problem in \eqref{main_opt} is to maximize the number of groups based on the communication matrix $B_X$. The first constraint is derived from $\Lambda_iB_X\leq (\sum_{j=1}^{N} \Lambda(i,j)-1)\mathbf{1}_M^T$ for each $i$, which ensures that the communicating databases in each group are unable to decode the stored messages. The third constraint avoids forming groups with a single database as we require security against each individual database, and the last constraint prevents a given database from being assigned to more than one group.
\end{remark}

Define the encoding function at the $n$th server $\Psi_n(\cdot)$ as, 
\begin{align}
    \Psi_n: \mathbb{F}_q ^{KL} \rightarrow \mathbb{F}_q^{\alpha_n}, \qquad n \in [N],
\end{align}
where $\alpha_n$ is the storage at the $n$th server. The user sends the queries to each server, such that the query sent to the $n$th database for the $k$th message is denoted by $Q_n^{[k]}$. Since the user has no prior information about the message contents, 
\begin{align}
    \label{no_info}I(W_{[K]};Q_n^{[k]}) = 0, \quad k \in [K].
\end{align}
In addition, the query sent to the $n$th server to retrieve $W_{\theta}$ must not convey any information about the message index, 
\begin{align}\label{privacy_req}
    I(\Theta ; Q_n^{[k]}) = 0, \quad k \in [K].
\end{align}
After receiving the queries, each database replies with a deterministic answer $\phi_n$, which is a function of the storage $\Psi_n(W_{[K]})$, 
\begin{align}
    \phi_n: \mathbb{F}_q ^{\alpha_n} \rightarrow \mathbb{F}_q ^{\beta_n},
\end{align}
where $\beta_n$ is the download cost for the $n$th server. Also,
\begin{align}\label{encoding}
    H(A_n^{[k]}|S_n,Q_n^{[k]}) = 0,
\end{align}
where $A_n^{[k]}$ is the answer from the $n$th server after receiving the query $Q_n^{[k]}$ for the $k$th message, and $S_n$ is the storage at the $n$th server.
After retrieving all the answers, the user reconstructs the required message using a decoding function, 
\begin{align}\label{decode}
    \mathcal{D}(k,A_1^{[k]}, \ldots, A_N^{[k]}): [K] \times \Pi_{n=1}^N \mathbb{F}_q^{\beta_n}  \rightarrow \mathbb{F}_q^L.
\end{align}
The required message is known by the user upon decoding, 
\begin{align}\label{final}
    H\left(W_{k}| \mathcal{D}(k,A_1^{[k]}, \ldots, A_N^{[k]}) \right)= H(W_{k}|k, A_{[N]}^{[\theta]}) =0.
\end{align}
For this setup, we define the average retrieval download cost per database as $\beta = \frac{1}{NR}$.

Since \eqref{main_opt} divides the servers into groups, $G_1, G_2,\ldots,G_g$, where $g$ is the number of groups, we define the download cost for a single group $G_i$, and the download cost for each database, respectively, as,
\begin{align}
    \beta_{G_i} = & \sum_{j \in G_i}H(A^{[k]}_j)/L,\\
    \beta_i =& H(A^{[k]}_i)/L, \quad i \in [N].
\end{align}
Similarly, we define the storage overhead per database, $\alpha$, as,
\begin{align}
    \alpha = \frac{\sum_{i=1}^{N} H(S_i)}{NKL}.
\end{align}
The storage per server, and the storage per group are,
\begin{align}
    \alpha_i=&\frac{1}{KL} H(S_i), \quad i \in [N],\\
    \alpha_{G_i} =& \sum_{k \in G_i} \alpha_k, \quad i \in [g],
\end{align}
 Finally, we define the region $\mathcal{R}$ as the region of the ordered pairs $(\alpha,\beta)$ that satisfy \eqref{no_info}, \eqref{privacy_req}, \eqref{encoding} and \eqref{final}.

\begin{remark}
    We note that the encoding of the messages and the retrieval scheme are done after grouping the servers using \eqref{main_opt} and the security scheme developed in \cite{asymmetric}.
\end{remark} 

\begin{remark}
    For each group of databases, we have the following trivial equality. If the databases $\{i_1,i_2, \ldots, i_k\}$ form a group after the optimization problem \eqref{main_opt}, then $\forall \ell \in [k]$,
    \begin{align}
        H(S_{G_i}^{\ell}|S_{G_i}^{1},\ldots,S_{G_i}^{\ell -1},S_{G_i}^{\ell+1},S_{G_i}^{k},W_{[1:K]})=0,
    \end{align}
    where $S_{G_i}^{m}$ is the storage at the $m$th server in the $i$th group.
\end{remark}

\begin{remark}
    Since one-time padding is used in the encoding scheme in \cite{asymmetric} to ensure information-theoretic privacy, we only focus on linear storage codes without loss of generality.
\end{remark}

\begin{remark}
    We note that both $\alpha$ and $\beta$ are in terms of the total number of servers $N$ not the number of effective servers \cite{asymmetric}. However, it can always be modified, without loss of generality, to the case of effective number of servers.
\end{remark}

\section{Main Results}\label{sec_main_results}

\begin{theorem}\label{special_case}
    For $N=4$ databases with communication matrix $i \leftrightarrow j$, $k \leftrightarrow \ell$, where $i,j,k,\ell \in [4]$, and $i\neq j \neq k \neq \ell$, with $K=2$ messages, any $(\alpha,\beta) \in \mathcal{R}$ must satisfy
    \begin{align}
        \label{special_case_eqs}
        \beta \geq \frac{3}{4}, \quad \alpha+2\beta \geq 2, \quad \alpha+6\beta \geq 3.
    \end{align}
\end{theorem}

\begin{remark}
    In comparison, without security constraints, i.e., PIR with $N=2$ servers and $K=2$ messages, with no security requirements, the storage-rate region is given by \cite{chao_storage},
    \begin{align}
          \beta \geq \frac{3}{4}, \quad \alpha+\beta \geq 2, \quad \alpha+6\beta \geq 6.
    \end{align}
\end{remark}

\begin{Proof}
To prove the first inequality, we notice that the optimal solution for the optimization problem in \eqref{main_opt} is to group the $k$th server with the $i$th server and the $j$th server with the $\ell$th server. Now, recall \cite[Theorem 1]{asymmetric}, which states that the scheme used achieves a rate equal to, 
\begin{align}\label{ach}
    R_{A-XSTPIR}= \frac{g}{\sum_{i=1}^g M_i} C_{TPIR}(T,g,K), 
\end{align}
where $C_{TPIR}(T,g,K)$ denotes the capacity of PIR with $g$ servers storing $K$ messages with $T$-colluding is given by,
\begin{align}
    C_{TPIR}(T,g,K)=\left(1+\frac{T}{g}+\dotsc+\left(\frac{T}{g}\right)^{K-1}\right)^{-1},
\end{align}
if the communication matrix satisfies
\begin{align}\label{feasibility}
     {N \choose X} - \Omega_X(B_X) \neq 0,
\end{align}
where $\Omega_i(\cdot)$ is defined as the number of columns in a matrix with weight $i$. Here, we have $g=2$, $M_1=M_2=2$, $K=2$ and $T=1$, and thus, the scheme achieves a rate $R =\frac{1}{3}$. 

Also, \cite[Theorem 2]{asymmetric} gives an upper bound on the rate as,
\begin{align}\label{ub}
    R_{A-XSTPIR} \leq \frac{\lambda}{M + \sum_{i=1}^M \zeta_{K,T}(\mathcal{X}_i)},
\end{align}
where $\lambda=\max_{n} \sum_{m=1}^M (1-B_X(n,m)) $, $M$ is the number of columns in $B_X$, $\mathcal{X}_i$ is the set of databases in the $i$th communication link, and $\zeta(\cdot)$ is the set function defined as,
\begin{align}\label{p(x)}
    \!\!\zeta_{K,T}(\mathcal{X}_i)\!=\! \frac{T}{N\!-\!|\mathcal{X}_i|} \!+\!\left(\!\frac{T}{N\!-\!|\mathcal{X}_i|}\!\right)^2\!+\! \cdots \!+\! \left(\!\frac{T}{N\!-\!|\mathcal{X}_i|}\!\right)^{K\!-\!1},
\end{align}
where $|\cdot|$ represents the cardinality of a set. In this setting, we have $\lambda = 1$, and $\zeta_{K,T}(\mathcal{X}_1) = \zeta_{K,T}(\mathcal{X}_2) = \frac{1}{2}$. Thus, $R \leq \frac{1}{3}$. Combining both results, we have $\beta \geq \frac{3}{4}$.

To prove the second inequality, recall the symmetry assumptions in \cite{multiround_pir} that can be used without loss of generality $\forall n, m \in [N]$, and $\forall j, k \in [K]$,
\begin{align}
    Q_n^{[k]}= Q_m^{[j]} \implies A_1^{[k]}=A_1^{[j]},  \text{ and}
\\
  H(A_n^{[k]}|\mathcal{Q})= H(A_m^{[j]}|\mathcal{Q}), \ H(S_n)=H(S_m) \nonumber\\
    \quad\implies \beta L \geq H(A_n^{[k]}), \ \alpha L \geq H(S_n),
\end{align}
where $\mathcal{Q} = \{Q_{G_1}^{[1]}, Q_{G_1}^{[2]}, Q_{G_2}^{[1]}, Q_{G_2}^{[2]}\}$. Then, we proceed as,
\begin{align}
    \alpha_{G_1}L + \beta_{G_2}L & \geq H(S_{G_1}) + H(A_{G_2}^{[1]})\\
    & \geq H(S_{G_1}) + H(A_{G_2}^{[1]}| Q_{G_2}^{[1]})\\
    & \geq H(S_{G_1},A_{G_2}^{[1]}| Q_{G_2}^{[1]})\\
    & = H(S_{G_1},A_{G_2}^{[1]},W_1| Q_{G_2}^{[1]})\\
    & = L + H(S_{G_1},A_{G_2}^{[1]}|W_1, Q_{G_2}^{[1]}) \\
    & = 2L +H(S_{G_1},A_{G_2}^{[3]}|W_1,W_2, Q_{G_2}^{[3]})\label{c1}\\
    & = 2L + \alpha_1L \label{ccc2},
\end{align}
where \eqref{c1} is due to \cite[(24)]{chao_storage}, and \eqref{ccc2} is due to the security constraint for each server.
Now, note that, $\alpha = \frac{1}{2}\alpha_{G_1}$,  $\beta = \frac{1}{2} \beta_{G_2}$, and $\alpha=\alpha_1$ which gives the second inequality.

To prove the third inequality, we find both upper and lower bounds for $I(A_{G_2}^{[1]}; A_{G_2}^{[2]}|\mathcal{Q})$. First, note that,
\begin{align}
    H(A_{G_1}^{[1]},A_{G_2}^{[1]}|W_1, \mathcal{Q}) & =  H(A_{G_1}^{[1]},A_{G_2}^{[1]},W_1| \mathcal{Q}) - H(W_1|\mathcal{Q})\\
    & = H(A_{G_1}^{[1]},A_{G_2}^{[1]}| \mathcal{Q}) - L \\
    & \leq 2 H(A_{G_1}^{[1]}) - L.
\end{align}
To find an upper bound for $I(A_{G_2}^{[1]}; A_{G_2}^{[2]}|\mathcal{Q})$, we proceed by using Ingleton's inequality,
\begin{align}
    I(A_{G_2}^{[1]}; &A_{G_2}^{[2]}|\mathcal{Q}) \nonumber \\
    \leq & I(A_{G_2}^{[1]}; A_{G_2}^{[2]}|W_1, \mathcal{Q}) +I(A_{G_2}^{[1]}; A_{G_2}^{[2]}|W_2, \mathcal{Q})\\
     =& 2 I(A_{G_2}^{[1]}; A_{G_2}^{[2]}|W_1, \mathcal{Q})\\
     =& 2 (H(A_{G_2}^{[1]}|W_1, \mathcal{Q}) + H(A_{G_2}^{[2]}|W_1, \mathcal{Q}) \nonumber \\  &-H(A_{G_2}^{[1]}, A_{G_2}^{[2]}|W_1, \mathcal{Q})) \label{36}\\
     =& 2(2H(A_{G_2}^{[1]}|W_1, \mathcal{Q}) - H(A_{G_2}^{[1]}, A_{G_2}^{[2]}|W_1 ,\mathcal{Q})) \label{37}\\
     \leq & 2(2H(A_{G_2}^{[1]}|W_1, \mathcal{Q}) + H(A_{G_1}^{[1]},A_{G_2}^{[1]}|W_1,\cq) \nonumber\\
     &- H(A_{G_1}^{[1]},A_{G_2}^{[1]},A_{G_2}^{[2]}|W_1,\cq) \nonumber \\ &- H(A_{G_2}^{[1]}|W_1,\cq))\\
     =& 2(H(A_{G_2}^{[1]}|W_1, \mathcal{Q}) + H(A_{G_1}^{[1]},A_{G_2}^{[1]}|W_1,\cq) \nonumber\\
     &-H(A_{G_1}^{[1]},A_{G_2}^{[1]},A_{G_2}^{[2]},W_2|W_1,\cq)\\
     \leq & 2 (2H(A_{G_1}^{[1]},A_{G_2}^{[1]}|W_1,\cq)- H(W_2))\\
     \leq & 2(2(2H(A_{G_1}^{[1]}|\cq)-L)-L) \\
     =& 8H(A_{G_1}^{[1]}|\cq) - 6L,
\end{align}
where \eqref{36} is due to conditioning reduces entropy, \eqref{37} is due to the symmetry assumption and the grouping technique in \cite{asymmetric}. For the lower bound, we proceed as follows, 
 \begin{align}
      I(A_{G_2}^{[1]};& A_{G_2}^{[2]}|\mathcal{Q}) \nonumber\\
      =&H(A_{G_2}^{[1]}|\cq)+H(A_{G_2}^{[2]}|\cq) - H(A_{G_2}^{[1]},A_{G_2}^{[2]}|\cq)\\
      \geq & 2 H(A_{G_1}^{[1]}|\cq) - H(A_{G_2}^{[1]},A_{G_2}^{[2]}|\cq)\\
      =& 2 H(A_{G_1}^{[1]}|\cq) - H(S_{G_2}|\cq)\\
      \geq &2 H(A_{G_1}^{[1]}|\cq) - \alpha_{G_2}L.
 \end{align}
Combining the upper and lower bounds gives the third inequality, completing the proof of Theorem \ref{special_case}.
\end{Proof}
\begin{remark}
    Unlike the no security case, the achievable region in Theorem \ref{special_case} can be further simplified only to the first two inequalities: $\beta \geq \frac{3}{4}, \ \alpha+2\beta \geq 2$, which is is shown in Fig.~\ref{fig_region}.
    \begin{figure}[h]
        \centering
        \includegraphics[width=0.6\columnwidth]{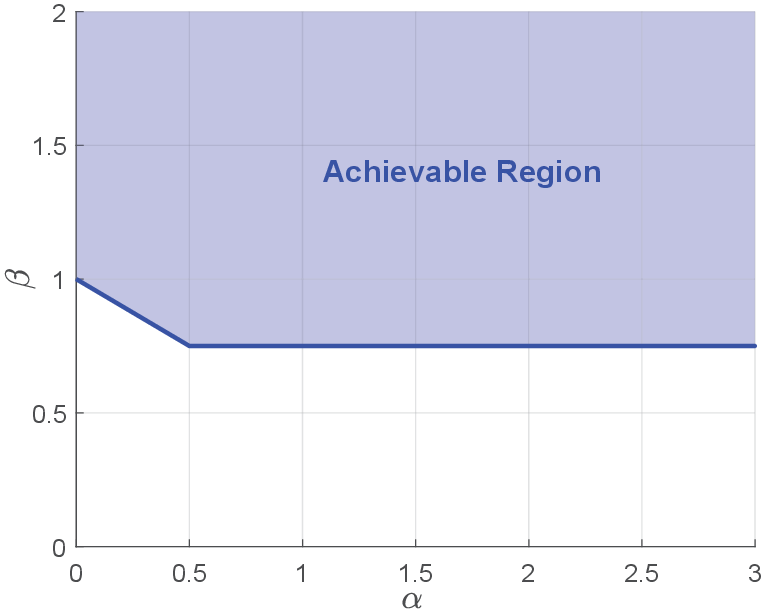}
        \caption{Achievable region for Theorem \ref{special_case}.}
        \label{fig_region}
    \end{figure}
\end{remark}
    
\begin{theorem}\label{general_case}
    For $g$ groups with $M_i$ databases in the $i$th group, and $K$ messages, any $(\alpha,\beta) \in \mathcal{R}$ must satisfy for all $\ell \in [g]$,  
    \begin{align}
    \Big( \!\!\sum_{i \in \{1,\ldots,g\} \atop i \neq \ell} \!\!\!M_i\Big)  \alpha &+ M_{\ell} \beta \geq  K\bigg(1 +\Big(\!\!\sum_{i \in \{1,\ldots,g\} \atop i \neq \ell} \!\!\!M_i\Big) -(g-1)\bigg),
\end{align}
 
\end{theorem}

\begin{Proof}
    We start with choosing any $\ell \in \{1,\ldots,g\}$
    \begin{align}
        &\sum_{i \in [g] \atop i \neq \ell}\alpha_{G_i} L + \beta_{G_{\ell}} L   \geq\sum_{i \in [g] \atop i \neq l}H(S_{G_i}) + H(A_{G_{\ell}}^{[1]})\\
        \geq & H(S_{G_1},\ldots,S_{G_{\ell-1}},S_{G_{\ell+1}},\ldots,S_{G_{g}}) + H(A_{G_{\ell}}^{[1]}| Q_{G_{\ell}}^{[1]})\\
        \geq & H(S_{G_1},\ldots,S_{G_{\ell-1}},S_{G_{\ell+1}},\ldots,S_{G_{g}},A_{G_{\ell}}^{[1]}| Q_{G_{\ell}}^{[1]})\\
        = & KL + H(S_{G_1},\ldots,S_{G_{\ell-1}},S_{G_{\ell+1}},\ldots,S_{G_{g}}|W_{1:K})\\
        = & K\big(1 +\!\!\sum_{i \in \{1,\ldots,g\} \atop i \neq \ell}\!\!\!M_i -(g-1)\big)L.
    \end{align}
    Now, note that, 
    \begin{align}
        \!\!\!\alpha = \frac{1}{M_1} \alpha_{G_1}, \ \beta = \frac{1}{M_g} \beta_{G_g}, \ \frac{\beta_{G_1}}{\beta_{G_i}} = \frac{M_1}{M_i}, \ \frac{\alpha_{G_1}}{\alpha_{G_i}} = \frac{M_1}{M_i}. \!
    \end{align}
    Thus,
    \begin{align}
        &\alpha_{G_1} \left(1+\frac{M_2}{M_1}+\ldots+\frac{M_{\ell-1}}{M_1} +\frac{M_{\ell+1}}{M_1}+ \ldots \frac{M_{g}}{M_1}\right) + M_g \beta \nonumber \\ & \geq K\bigg(1 +\!\!\sum_{i \in \{1,\ldots,g\} \atop i \neq \ell}\!\!\!M_i -(g-1)\bigg),
    \end{align}
    completing the proof of Theorem \ref{general_case}.    
\end{Proof}

\begin{corollary}
     For $g$ groups with $M_i$ databases in the $i$th group, $\sum_i M_i= N $ and $K$ messages,  any $(\alpha,\beta) \in \mathcal{R}$ must satisfy 
    \begin{align}
        (N - M_{\ell}) \alpha + M_{\ell} \beta \geq K\left(1 +(N-M_{\ell}) -(g-1)\right),
    \end{align}
    for all $l \in \{1,\ldots, g\}$.   
\end{corollary}

\begin{corollary}\label{uniform_grouping}
    For $g$ groups with $d$ databases per group, $K$ messages, and $gd=N$, any $(\alpha,\beta) \in \mathcal{R}$ must satisfy
    \begin{align}
        d (g-1)\alpha + d\beta \geq K \left(1+(d-1)(g-1)\right).
    \end{align}
\end{corollary}

\section{Illustrative Example}\label{sec_example}
Assume that we have $N=4$ servers, $T=1$ colluding parameter (i.e., no collusion), $K=2$ messages, and all single-way communications need to be secure, i.e., each server cannot know any information about the messages. In addition, assume that server 1 communicates with server 2, denoted by $ 1 \leftrightarrow 2$, and server 3 communicates with sever 4, denoted by $ 3 \leftrightarrow 4$. Using the grouping method and the optimization problem in \eqref{main_opt}, the optimal grouping is server 1 with server 3, and server 2 with server 4. Now, let $L=g^K =2^2 =4$. For the storage in the first group pair, we have,
\begin{align}
    S_1 =& \big(a_1+N_1(1,1), a_3+N_1(1,3), b_1+N_1(2,1),\nonumber \\
    & \ b_3+N_1(2,3), a_2+b_2+N_1(1,2)+N_1(2,2), \nonumber \\
    & \ a_4+b_4+N_1(1,4)+N_1(2,4)\big),\\
    S_3 =& \big(N_1(1,1), N_1(1,3),N_1(2,1), N_1(2,3), \nonumber \\ & \ N_1(1,2)+N_1(2,2), N_1(1,4)+N_1(2,4)\big).
\end{align}
Similarly, for the second group, we have, 
\begin{align}
     S_2 =& \big(a_2+N_2(1,2), a_4+N_2(1,4),b_2+N_2(2,2), \nonumber \\ 
     & \ b_4+N_2(2,4), a_1+b_3+N_2(1,1)+N_2(2,3), \nonumber \\ 
     & \ a_3+b_1+N_2(1,4)+N_2(2,3)\big),\\
     S_4 =& \big(N_2(1,2), N_2(1,4),N_2(2,2), N_2(2,4), \nonumber \\ 
     & \ N_2(1,1)+N_2(2,3), N_2(1,4)+N_2(2,3)\big). 
\end{align}
Tables \ref{ret_1_table} and \ref{ret_2_table} show the retrieval process that can be used by the user with probability $\frac{1}{2}$, each.
\begin{table}[t]
    \centering
    \caption{Retrieval scheme with probability $\frac{1}{2}$.}
    \begin{tabular}{|p{0.7cm}|p{3.25 cm}|p{3.25 cm}|}
        \hline
    & Want $W_1$& Want $W_2$\\
    \hline
    DB1& $\begin{aligned}
        &a_1+N_1(1,1),\\
    &b_1+N_1(2,1),\\
    &a_2+b_2+N_1(1,2)\\
    & +N_1(2,2)
    \end{aligned}$
     & $\begin{aligned}
         &a_1+N_1(1,1),\\
         &b_1+N_1(2,1),\\
         &a_2+b_2+N_1(1,2)\\ &+N_1(2,2)
     \end{aligned}$\\
    \hline
    DB2 & 
    $\begin{aligned}
        &a_4+N_2(1,4), \\ 
        &b_2+N_2(2,2),\\
    &a_3+b_1+N_2(1,3) \\ 
    &+N_2(2,1)
    \end{aligned}$ &
    $\begin{aligned}
        &a_2+N_2(1,2),\\
        &b_4+N_2(2,4),\\
    &a_1+b_3+N_2(1,1)\\ 
    &+N_2(2,3)
    \end{aligned}$\\
    \hline
    DB3& 
   $\begin{aligned}
        &N_1(1,1), \quad N_1(2,1),\\
        & N_1(1,2)+N_1(2,2)
    \end{aligned}$ & $\begin{aligned}
        &N_1(1,1), \quad N_1(2,1),\\
        & N_1(1,2)+N_1(2,2)
    \end{aligned}$\\
    \hline
    DB4 & 
    $\begin{aligned}
        &N_2(1,4), \quad N_2(2,2),\\
        &N_2(1,3)+N_2(2,1)
    \end{aligned}$ & 
    $\begin{aligned}
     &N_2(1,2), \quad N_2(2,4),\\
     &N_2(1,1)+N_2(2,3)
    \end{aligned}$\\
    \hline
    \end{tabular}
    \label{ret_1_table}
\end{table}


\begin{table}[t]
    \centering
    \caption{Retrieval scheme with probability $\frac{1}{2}$.}
    \begin{tabular}{|p{0.7cm}|p{3.25 cm}|p{3.25 cm}|}
        \hline
    & Want $W_1$& Want $W_2$\\
    \hline
    DB1& 
    $\begin{aligned}
      &a_3+N_1(1,3),\\
      &b_3+N_1(2,3),\\
      &a_4+b_4+N_1(1,4)\\
      &+N_1(2,4)
    \end{aligned}$ & $\begin{aligned}
      &a_3+N_1(1,3),\\
      &b_3+N_1(2,3),\\
      &a_4+b_4+N_1(1,4)\\
      &+N_1(2,4)
    \end{aligned}$ \\
    \hline
    DB2 & $\begin{aligned}
        &a_2+N_2(1,2),\\
        &b_4+N_2(2,4),\\
    &a_1+b_3+N_2(1,1)\\ 
    &+N_2(2,3)
    \end{aligned}$& $\begin{aligned}
        &a_4+N_2(1,4), \\ 
        &b_2+N_2(2,2),\\
    &a_3+b_1+N_2(1,3) \\ 
    &+N_2(2,1)
    \end{aligned}$ \\
    \hline
    DB3& 
    $\begin{aligned}
        &N_1(1,3), \quad N_1(2,3)\\
        &N_1(1,4)+N_1(2,4)
    \end{aligned}$&
    $\begin{aligned}
        &N_1(1,3), \quad N_1(2,3),\\
        &N_1(1,4)+N_1(2,4)
    \end{aligned}$ \\
    \hline
    DB4 &
    $\begin{aligned}
        &N_2(1,2), \quad N_2(2,4),\\
        &N_2(1,1)+N_2(2,3)
    \end{aligned}$
     & $\begin{aligned}
         &N_2(1,4), \quad N_2(2,2), \\
         &N_2(1,3)+N_2(2,3)
     \end{aligned}$\\
    \hline
    \end{tabular}
    \label{ret_2_table}
\end{table}

It is clear that the rate of this scheme is $R = \frac{4}{3\times 4} = \frac{1}{3}$, which is optimal for the case of the fully replicated storage. However, here instead of storing $8$ symbols per server, we store only $6$ symbols per server with a storage efficiency $\frac{6}{8} = \frac{3}{4}$. 

To prove the privacy of the scheme, we check the dependence between the required message index and the queries sent to each individual server. We note that, we show the privacy proofs only for servers $1$ and $2$, since the privacy proof for servers $3$ and $4$ follow the same lines. We drop the noise terms in the following calculations for simplicity. For server 1,
\begin{align}
    &\mathbb{P} \left[  \Theta = 1 | Q_1 = (a_1,b_1,a_2+b_2)\right] \nonumber \\
    &\quad = \mathbb{P}\left[ \Theta = 2| Q_1 = (a_1,b_1,a_2+b_2)\right]=\frac{1}{2}.
\end{align}
For $Q_1 =\left( a_3, b_3, a_4+b_4 \right)$, the proof follows the same lines. In addition, for server 2, we have,
\begin{align}
    &\mathbb{P} \left[ \Theta \!=\! 1| Q_2 \!=\! (a_4,b_2,a_3\!+\!b_1)\right] \nonumber\\
    &= \mathbb{P}[\text{scheme}\!=\!1] \mathbb{P} \left[ \Theta \!=\!1| \text{scheme}\!=\!1, Q_2 \!=\! (a_4,b_2,a_3\!+\!b_1)\right] \nonumber\\
    & \ \ \ + \mathbb{P}[\text{scheme}\!=\!2] \mathbb{P} \left[  \Theta \!=\! 1|\text{scheme}\!=\!2, Q_2 \!=\! (a_4,b_2,a_3\!+\!b_1)\right] \nonumber \\
    &  = \frac{1}{2}+0 =\frac{1}{2}.
\end{align}
For $Q_2 =\left( a_2, b_4, a_1+b_3 \right)$, the proof follows the same lines.

We note that even if servers $1$ and $3$ are colluding, and servers $2$ and $4$ are colluding, the rate remains the same. This shows an interesting connection between the grouping approach and the arbitrary collusion among the servers within a group. Specifically, the servers in the same group can collude without affecting the privacy, and the rate remains the same. 

The following theorem provides the first capacity result that connects arbitrary collusion and communication based on the previous observation.

\begin{theorem}
    For the case of $N=4$ servers and $K=2$ messages with the following set of communicating links, $\mathcal{X} = \{\{1,2\}, \{3,4\}\}$, and the following set of colluding servers $\mathcal{T}= \{\{1,3\},\{2,4\}\}$, the capacity is given by,
    \begin{align}
        C(N,K,\mathcal{X},\mathcal{T}) = \frac{1}{3}.
    \end{align}
    Further, the optimal message lengths are equal to $4$, and a total of $6$ symbols are stored in each server.
\end{theorem}

More generally, let $\lambda = \max_{n}\sum_{m=1}^M\left(1-B_X(n,m) \right)$, and $|\mathcal{X}_i| = N-g$, $i \in [M]$ where $g$ is the number of groups upon using \eqref{main_opt}. Define $\eta_K(\mathcal{X}_i) = \zeta_{K,1}(\mathcal{X}_i)$, then we have the capacity result given in the following theorem.

\begin{theorem} \label{thm4}
    Let $N$ and $K$ be the number of servers and the number of messages, respectively. Let $B_X$ be the communication matrix with $M$ communications links. If $\frac{\lambda}{M} = \frac{g}{\sum_{i}M_i}$, where $g$ is the number of groups and $M_i$ is the number of servers in the $i$th group after solving \eqref{main_opt}, and $|\mathcal{X}_i| = N-g$. In addition, let $\mathcal{T}$ contain the sets of servers in each group. Then, the capacity is given by,
    \begin{align}
        C(N,K,\mathcal{X},\mathcal{T}) =& \frac{\lambda}{M\left(1 + \eta_K(\mathcal{X}_1)\right)}\\
        =& \frac{\lambda}{M}C_{PIR}(N-|\mathcal{X}_1|,K) \\
        =& \frac{\lambda}{M} C_{PIR}(g,K).
    \end{align}
\end{theorem}

\begin{remark}
    The result in Theorem \ref{thm4} is remarkable since the asymmetric collusion in this case does not affect the capacity in some settings for asymmetric security.     
\end{remark}

As a numerical example, if $1 \leftrightarrow 2 \leftrightarrow 3$, and $4 \leftrightarrow 5 \leftrightarrow 6$, with $\mathcal{T} =\{\{1,4\}, \{2,5\}, \{3,6\}\}$ we have $g=3$ after \eqref{main_opt}, and $|\mathcal{X}_1|=|\mathcal{X}_2| =3 = 6-3$. Thus, the capacity is $C = \frac{1}{2}C_{PIR}(3,K)$. It is also worth noting that if other extra links exist, e.g., $1 \leftrightarrow 5$, or $1 \leftrightarrow 2 \leftrightarrow 6$, the capacity remains the same. Thus, there might exist some relaxations that can be done on Theorem~\ref{thm4}.

\section{Conclusion}
In this paper, we considered the storage-rate trade-off for the asymmetric XPIR (A-XPIR) problem, with a given fixed communication matrix for the databases. We derived trade-offs for the general problem, and analyzed the $N=4$ servers with two communication links case in depth as a separate result. In addition, we derived capacity results for certain special families of arbitrary collusion and communication settings. 

\newpage
\bibliographystyle{unsrt}
\bibliography{references.bib}

\begin{thebibliography}{10}

\bibitem{chor}
B.~Chor, E.~Kushilevitz, O.~Goldreich, and M.~Sudan.
\newblock Private information retrieval.
\newblock {\em Jour. of the ACM}, 45(6):965--981, November 1998.

\bibitem{c_pir}
H.~Sun and S.~Jafar.
\newblock The capacity of private information retrieval.
\newblock {\em IEEE Trans. Info. Theory}, 63(7):4075--4088, July 2017.

\bibitem{colluding}
H.~Sun and S.~Jafar.
\newblock The capacity of robust private information retrieval with colluding databases.
\newblock {\em IEEE Trans. Info. Theory}, 64(4):2361--2370, April 2018.

\bibitem{arbitrarycollusion}
X.~Yao, N.~Liu, and W.~Kang.
\newblock The capacity of private information retrieval under arbitrary collusion patterns for replicated databases.
\newblock {\em IEEE Trans. Info. Theory}, 67(10):6841--6855, July 2021.

\bibitem{sun_eaves}
Q.~Wang, H.~Sun, and M.~Skoglund.
\newblock The capacity of private information retrieval with eavesdroppers.
\newblock {\em IEEE Trans. Info. Theory}, 65(5):3198--3214, December 2018.

\bibitem{nan_eaves}
J.~Cheng, N.~Liu, W.~Kang, and Y.~Li.
\newblock The capacity of symmetric private information retrieval under arbitrary collusion and eavesdropping patterns.
\newblock {\em IEEE Trans. Info. Foren. Security}, 17:3037--3050, August 2022.

\bibitem{csa}
Z.~Jia, H.~Sun, and S.~Jafar.
\newblock Cross subspace alignment and the asymptotic capacity of {$X$}-secure {$T$}-private information retrieval.
\newblock {\em IEEE Trans. Info. Theory}, 65(9):5783--5798, May 2019.

\bibitem{asymmetric}
M.~Nomeir, S.~Vithana, and S.~Ulukus.
\newblock Asymmetric {X}-secure {T}-private information retrieval: More databases is not always better.
\newblock In {\em CISS}, March 2024.

\bibitem{grpahbased_pir}
N.~Raviv, I.~Tamo, and E.~Yaakobi.
\newblock Private information retrieval in graph-based replication systems.
\newblock {\em IEEE Trans. Info. Theory}, 66(6):3590--3602, November 2019.

\bibitem{shreya_graph_1}
S.~Meel and S.~Ulukus.
\newblock Effect of full common randomness replication in symmetric {PIR} on graph-based replicated systems.
\newblock Available online at arXiv:2510.25736.

\bibitem{shreya_graph_2}
S.~Meel and S.~Ulukus.
\newblock Symmetric private information retrieval {SPIR} on graph-based replicated systems.
\newblock In {\em IEEE Globecom}, December 2025.

\bibitem{shreya_graph_3}
X.~Kong, S.~Meel, T.~Maranzatto, I., and S.~Ulukus.
\newblock New capacity bounds for {PIR} on graph and multigraph-based replicated storage.
\newblock {\em IEEE Trans. Info. Theory}, 72(1):691--709, January 2026.

\bibitem{shreya_graph_4}
S.~Meel, X.~Kong, T.~Maranzatto, I.~Tamo, and S.~Ulukus.
\newblock Private information retrieval on multigraph-based replicated storage.
\newblock In {\em IEEE ISIT}, June 2025.

\bibitem{JAFAR_STAR_4}
Y.~Yao and S.~Jafar.
\newblock The capacity of 4-star-graph {PIR}.
\newblock In {\em IEEE ISIT}, June 2023.

\bibitem{karim_graph}
K.~Banawan and S.~Ulukus.
\newblock Private information retrieval from non-replicated databases.
\newblock In {\em IEEE ISIT}, July 2019.

\bibitem{jafar_asymptotic_graph}
Z.~Jia and S.~Jafar.
\newblock On the asymptotic capacity of {X}-secure t-private information retrieval with graph-based replicated storage.
\newblock {\em IEEE Trans. on Info. Theory}, 66(10):6280--6296, July 2020.

\bibitem{chao_storage}
C.~Tian.
\newblock On the storage cost of private information retrieval.
\newblock {\em IEEE Transactions on Information Theory}, 66(12):7539--7549, December 2020.

\bibitem{sun-chao-storage}
C.~Tian, H.~Sun, and J.~Chen.
\newblock A {S}hannon-theoretic approach to the storage-retrieval tradeoff in {PIR} systems.
\newblock In {\em IEEE ISIT}, August 2018.

\bibitem{multiround_pir}
H.~Sun and S.~Jafar.
\newblock Multiround private information retrieval: Capacity and storage overhead.
\newblock {\em IEEE Trans. Info. Theory}, 64(8):5743--5754, January 2018.

\end{thebibliography}
\end{document}